\documentclass{elsart}
\usepackage{amssymb}
\usepackage{graphicx}
 
\input{tcilatex}

\begin{document}   
\begin{frontmatter}

\title{Wang-Landau study of the critical behavior of the bimodal 3D-Random
Field Ising Model}    
\author[CERGY]{Laura Hern\'andez\thanksref{mail1}}
\author[CNEA]{Horacio Ceva\thanksref{mail2}}
\address[CERGY]{Laboratoire de Physique Th\'eorique et Mod\'elisation, UMR CNRS-Universit\'e de Cergy Pontoise,
2 Av Adolphe Chauvin, 95302 Cergy-Pontoise Cedex, France}
\address [CNEA]{Departamento de F{\'{\i}}sica, Comisi{\'o}n Nacional de Energ{\'\i }a At{\'o}mica, Avenida del Libertador 8250, 1429 Buenos Aires, Argentina}   
\thanks[mail1]{Laura.Hernandez@u-cergy.fr}
\thanks[mail2]{ceva@cnea.gov.ar}

\begin{abstract}
We apply the Wang-Landau method to the study of the critical behavior of
the three dimensional Random Field Ising Model with a bimodal probability
distribution. For high values of the random field intensity we find that the energy probability distribution at the transition temperature is double peaked, suggesting that the phase transition is first order. On the other
hand, the transition looks continuous for low values of the field intensity.
In spite of the large sample to sample fluctuations observed, the double
peak in the probability distribution is always present for high fields.
\end{abstract}


\end{frontmatter}

\section{Introduction}

Out of the many disordered systems that have given place to lively arguments
about central aspects of their behavior, the 3D Random Field Ising Model
(RFIM)  has been the subject of many publications along the
years. Given the fact that until now it has proved impossible to find exact
results, all kind of numerical techniques and solutions have been
implemented, mostly with a high degree of sophistication. Every time that a
new technical  breakthrough arises, there is an avalanche of new works in
this field. The last example of this is the proliferation of ``extended
 ensemble simulations'' methods. Monte Carlo simulations works done for the
last decades used to  sample the phase space with a canonical
ensemble and suffered from  two kind of well known problems:
critical slowing down near the critical temperature of second order
transitions on one hand, and on the other, meta-stable states
trapping when the free energy landscape has several minima, and high
potential barriers must be overcome to pass from one local minimum
to another. Hence it is difficult to simulate systems undergoing a
first order transition because
 the system may remain in one potential well and the other may not be observed.
The common characteristics of ``extended ensemble methods'', also generally called
``multicanonical methods'', is that the phase state of the system is not
sampled using   the canonical ensemble~\cite{multican}. In particular, the
Wang-Landau  method~\cite{wl1}  builds the density of states $(DOS)$ of the
system iteratively, by performing a random  walk in the
energy space. As  the roughness of the energy landscape is not
 an obstacle for the evolution of the simulation any more, this method is well
 adapted to the study of disordered systems which have been particularly difficult
 to simulate so far.

We believe that the central point that is still a matter of controversy, is
the identification of the kind of  phase transition (first order or
continuous)\ of the 3D RFIM, specially in the bimodal variety, i.e. when the
random fields can only have the values $\pm h$. In this sense the WL method
is a powerful tool to study systems undergoing first-order (as well as
continuous) phase transitions. Detailed studies of its applications to well
known systems presenting first-order transitions exists~\cite{wl2} and many
variations have been proposed to speed-up the algorithm or to reduce errors~
\cite{zhou-blatt}-\cite{zhou2}.

It is not our intention to give an exhaustive description of all the types
of results published in the literature (see, for instance, \cite{ref1}-
\cite{ref14}).
We merely point out  that a great majority of the works  are addressed to
the Gaussian distribution of random fields, although there also exists a
conspicuous group of works using the bimodal distribution.
However, almost all these works assume from the start that ``nowadays, it is
widely believed that  the phase transition is of second-order''. Hence, they
are centrally oriented to calculate critical indices and eventually try to
fit them into relationships which  are valid only if the transition is
continuous. To the best of our knowledge, there is not yet a work with a
good fit in this sense. In some works the calculated critical exponents are field dependent as in \cite{ref11}.  Other authors propose a modification to scaling introducing an extra exponent \cite{DFisher} \cite{jolicoeur}.
 In other works it is found that the critical index $\beta $ is almost (or even exactly) zero,
implying a discontinuity of the order parameter at the critical
temperature\cite{ref3}\cite{ref11a}. For instance, in Ref.\cite{ref3} the authors estimate that
in order to be able to see that the order parameter goes continuously to zero
one needs to use ``astronomically large'' samples (with $10^{21}$ or more spins).
In this particular case they decided not to reach to conclusions from the
study of the magnetization; instead, they made a rather elaborated study of
the stiffness of the domain walls present in the simulations to conclude
that the transition is continuous.

Hence we believe that the question concerning the order of the transition is still open.
In this work we will address this point, by calculating the energy
probability distribution $P_T(E)$, at the transition temperature. This is done
starting from a rather detailed calculation of $DOS$%
, obtained with the WL\ method.  From this point of view, the question is
rather clear cut:\ if $P_T(E)$ at the critical temperature has a single peak,
it is a continuous transition; if it has two peaks, the transition is first
order.

There are many works related with specific technical aspects of the WL method.
 Nevertheless still many aspects of the method have to be
tailored to adapt it to a particular system, specially in the case
of disordered systems where less studies
exist\cite{ref13}~\cite{ref14}~\cite{sg1}. Therefore, in the
following we explain some details of the application of the WL
method to the 3D-RFIM, and afterwards present our results and
conclusions. We anticipate that our results indicate that the
transition is first order.

\section{Description of the model and analysis of the applied method}

\label{description}

We study by the WL method the 3D RFIM in a cubic lattice of linear size $L$,
with a bimodal  distribution of the random field. Hence the probability
distribution of the random  fields reads:

\begin{equation}
p(h_{i})=\frac{1}{2}\left[ \delta \left( h_{i}-h_{0}\right) +\delta \left(
h_{i}+h_{0}\right) \right]   \label{prob}
\end{equation}

where $h_0$ is the intensity of the random field. In this work, the value of
the nearest neighbour interaction of the Ising lattice is set to $J=1$.

The WL method is an iterative procedure that allows for the calculation of the DOS $g(E)$ by performing a random walk in the energy space. Knowing $g(E)$ one can calculate the probability of finding a given
energy, $P_T(E)$, and the thermodynamic average of any energy dependent quantity $A(E)$ as:

\begin{equation}
P_T(E)=\frac{g(E) exp(-\beta E)}{\sum_{E}g(E)exp(-\beta E)}
\end{equation}

\begin{equation}
<A>_T= \frac{\sum_{E}g(E)A(E)exp(-\beta E)}{\sum_{E}g(E)exp(-\beta E)}%
=\sum_{E}P_T(E)A(E)
\end{equation}

where, as usual, $\beta=1/T$.

   A  well known way to
speed up the algorithm is to split the whole energy interval in sub-intervals
where the WL algorithm converges faster. Then the different portions of the
DOS corresponding to each sub-interval must be joined to obtain the DOS in
the whole energy range. This is the multi-range version of the WL
algorithm ~\cite{wl2}.

 We have found that, in the case of the RFIM, this multi-range version of the
 WL method is not only  useful to speed up the algorithm but essential to
 obtain its convergence.  In fact convergence is not achieved when performing
 a single-range run covering the whole energy interval for linear sizes
 $L \ge 10$.

Hence in our case the pertinent energy per site interval for the calculation of
averages, [-3,0.3] in units of $J$, is divided into successive overlapping
sub-intervals  (one should notice that higher energies will give a very low
contribution to the average sums). The convergence of the algorithm, indicated
by the flatness of the histogram for values of the WL factor f close to 1, depends on
the explored region of the energy domain. We can roughly identify  three regions:
 a region near the ground state (I), a middle energy region (II) and a high
energy region (III) where the system is disordered.

For a given flatness criterion, convergence is hard to reach in the
first two regions. In region I, this is so due to the presence of
the ground state. Region II is associated to the energy interval
where the peak (or peaks) of $P_{T^*}(E)$, the probability
distribution at the transition temperature, is (are) located. In the
third region convergence is generally easily achieved. The exact
location of these regions depends on the value $h_0$.

We overcame this convergence problem by reducing the width of the considered
energy sub-intervals,  in a substantial part of regions I and II.
On the other hand, in the third region, convergence is achieved even
when studying quite large sub-intervals.

The estimate DOS in a given sub-interval is considered accurate
when the energy histogram is flat in that sub-interval. We
succeeded to greatly improve the flatness of the histograms by re-running
\textit{the whole WL algorithm iteratively}.

We proceeded as follows: as a first step the WL algorithm was run with some
flatness criterion, getting an estimate of the density of states $g(E)$.
Typically we considered that convergence is achieved when 90\% (80\% for the
bigger sizes) of the histogram points are found within an interval of 10\% (20\%)
of the theoretical energy histogram.

Then this first estimate of $g(E)$ was used as an input for a new complete WL run. We refer
 to this as a \textit {second iteration step}. Moreover we use a much  more
 restrictive flatness criterion and  a relatively low initial
value of the WL factor (typically $ln(f)=10^{-5}$). At this stage the convergence
criterion was improved:  99\% of the histogram points should be
within 1\% of the theoretical value of the histogram.  We also impose a
maximum number of Monte Carlo steps per spin (MCS/s) to prevent the program
from running forever if the flatness criterion is not matched.
This  very strong flatness criterion  forces the algorithm  to reach the maximum
MCS/s allowing for  better statistics.

In most of the cases this produces extremely flat histograms except for the
borderline energies of the sub-intervals, specially in region II.

\begin{figure}[tbp]
\begin{center}
\includegraphics[width=9.5cm]{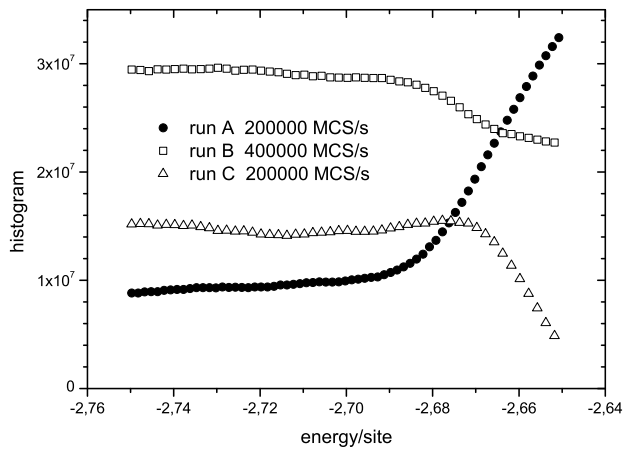}

\caption{Convergence study in zone II. $L=24$, $h_0=2.1$,  runs A
and B are a second step  iteration of the whole  WL algorithm,  run
C is a third step. Doubling the maximum MCS/s number causes an
important improvement in the flatness of the histogram (compare
histogram A and B). However, it is more efficient to take the input
of run A and re-run the WL algorithm (run C) (for the same total
amount MCS/s). The region where the histogram is almost completely
flat increases beyond that corresponding to run B. For clarity only
some points of the histograms are plotted.} \label{convergence_a}
\end{center}
\end{figure}

Figure~\ref{convergence_a} shows histograms obtained at different iteration
steps for the hardest  energy interval (with respect to convergence of the algorithm) in region II.  Different runs are shown.

Run A  has been performed with  a maximum of 200000 MCS/s  for each value of the WL factor
and run B with 400000 MCS/s. The energy range where the energy histogram is flat is wider
for the longer run, but it still doesn't cover the whole sub-interval.

A  better result is obtained with run C where the almost perfect
flat region is extended beyond the one of run B.  Run C has  a
maximum of 200000 MCS/s for each f-value but it is  \textit{a third
iteration step}: the initial guess of the density of states is the
$g(E)$ obtained  from run A. So we see that for a given total amount
of MCS/s, it is more efficient to perform two iteration steps of the
whole WL algorithm with 200000 MCS/s each, than only one with 400000
MCS/s.

\begin{figure}[tbp]
\begin{center}
\includegraphics[width=9.5cm] {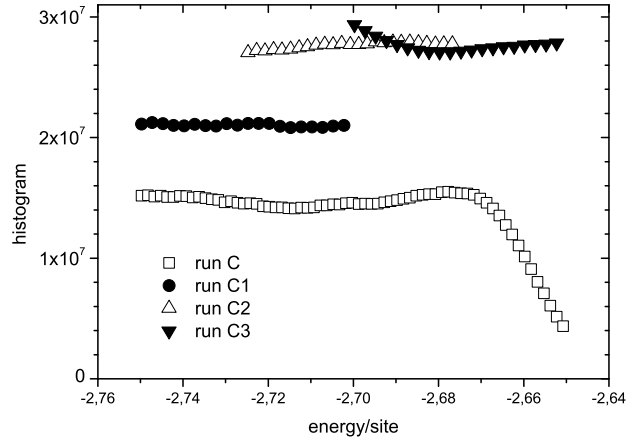}

\caption{Convergence study in zone II. $L=24$, $h_0=2.1$, Here all
the runs are a third iteration step (this means that the input was
taken from run A). Comparison between  the results issued from run C
over the whole sub-interval  and those obtained splitting the
interval into three overlapping ones (C1,C2,C3).  For clarity only
some points of the histograms are plotted.} \label{convergence_b}
\end{center}
\end{figure}

To further improve this convergence, we  combine this third iteration step with an extra
splitting of the energy subinterval in region II.   In Figure~\ref{convergence_b} we
compare the same run C already shown in  figure~\ref{convergence_a}  with three new
runs  performed in smaller overlapping energy intervals. The three of them are a third
iteration steps with the same initial guess that run C. In this way we achieve convergence
in region II.

As usual, the global density of states $g(E) , \forall E$ was built up by joining
the different parts obtained in each interval.

\section{Results}

We have studied 3D lattices of linear size $L$=10, 16, 20, 24, 30. For a given size and $h_0$ value, different realizations (around 10) of the quenched disorder have been studied, each one of them is called a ``sample'' in the following. For some of these samples we
have studied the behavior of the system as a function of $h_0$.

Our results show two clearly distinct critical behaviors according to the
value of $h_0$.

\begin{figure}[tbp]
\begin{center}
\includegraphics [width=9.5cm]{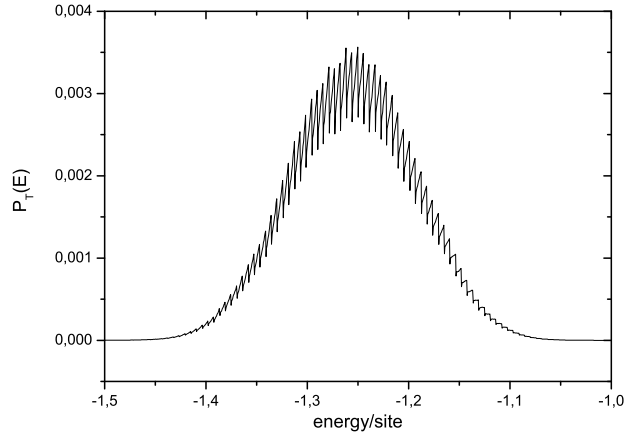}
\caption{$L=24$, $h_0=0.5$  Single-peaked probability density at the
critical temperature $T_c=4.34$}
\label{prob_low_h}
\end{center}
\end{figure}

For low $h_0$ values the specific heat curve as a function of temperature, $C(T)$, shows the characteristic peak of a second order transition in a
finite system at a certain critical temperature $P_{T_c(L)}(E)$ (see figure~\ref%
{sh_low_h}). The corresponding probability density at $T_c(L)$ is shown in
Figure~\ref{prob_low_h}. Only one peak is observed, indicating that the
system is dominated by thermal fluctuations while the presence of the random
field is only a small perturbation.

\begin{figure}[tbp]
\begin{center}
\includegraphics [width=9.5cm] {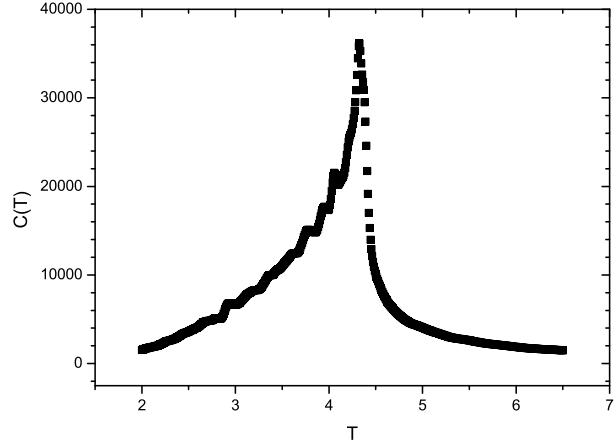}
\caption{$L=24$, $h_0=0.5$ Specific heat as a function of temperature.}
 \label{sh_low_h}
 \end{center}
 \end{figure}

\begin{figure} [tbp]
\begin{center}
\includegraphics [width=9.5cm]{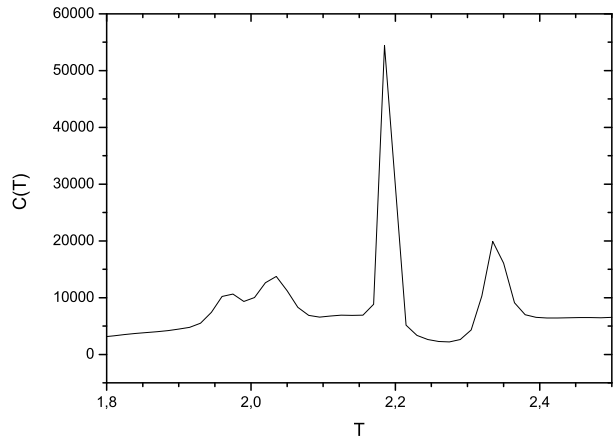}
\caption{$L=24$, $h_0=2.1$ Specific heat as a function of temperature. Multiple
peaks appear, their locations and heights are sample-dependent.}
\label{sh_high_h}
\end{center}
\end{figure}

On the other hand, for higher fields the $C(T)$ curves may have more
than one peak with, in general, a dominant peak, and a few secondary ones, as can be seen in figure~\ref{sh_high_h}.

Moreover, the position of these $C(T)$ peaks is sample-dependent,
suggesting that  \textit{each sample has its own transition
temperature $T^*$}. The consequence of these large sample to sample fluctuations
is to prevent from performing  a naive averaging over the quenched disorder.

The analysis of the $P_T(E)$ curves near the temperatures where the maxima of $C(T)$ are located provides an interpretation for this  multiplicity of peaks.

\begin{figure} [tbp]
\begin{center}
\includegraphics [width=9.5cm]{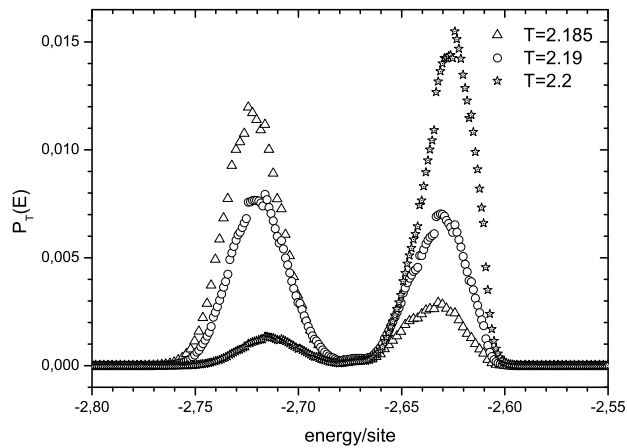}
\caption{$L=24$, $h_0=2.1$  Double-peaked probability density at three
 temperatures in the transition region.  It can be seen how the peak is
 shifted from low to high energy values as the temperature increases.  The
 transition temperature may be estimated from these curves as the value
 where the two peaks have the same height (here $T=2.19$). For the sake of clarity only some points of the curves are plotted.}
\label{prob_high_h}
\end{center}
\end{figure}

In figure~\ref{prob_high_h} we show the corresponding $P_T(E)$
curves for three different temperatures in the region of the highest
peak  of the specific heat of figure~\ref{sh_high_h}. 
A double peak structure clearly appears, showing the coexistence of states with different energies separated by an energy  region where the probability is zero. The transition temperature $T^*$  may be  determined by  the temperature  where the two peaks in $P_{T^*}(E)$ have equal height. Here it is $T^*\approx 2.19$.
A qualitative study of the size effect shows that the double peak in                                                                 $P_{T^*}(E)$ is present for $L  \geq  16$, and enhances with size.

\begin{figure}[tbp]
\begin{center}
\includegraphics [width=9.5cm,clip]{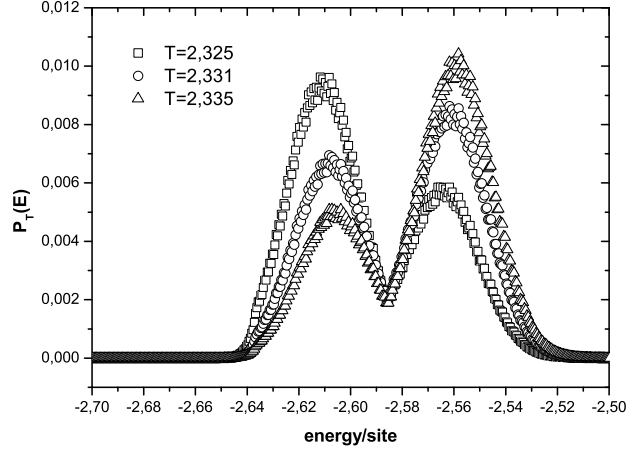}
\caption{$L=24$, $h_0=2.1$  Probability density at three
 temperatures around a secondary peak of $C(T)$. There is a
 coexistence of energies with non zero probability, indicating the
 existence of meta-stable states separated by energy barriers of
 different heights. For the sake of clarity only some points of
the curves are plotted.} \label{prob_sec_peak}
\end{center}
\end{figure}

The shape of the $P_T(E)$ corresponding to the secondary peaks of the $C(T)$ is also far from the one characterizing a second order
transition (see figures ~\ref{prob_low_h} and \ref{prob_sec_peak}). It shows a structured peak, somehow similar to a double peak, but it should be noticed that the minimum in between the maxima does not go to zero.  
 The same structure is found for the other secondary peaks.

This allows for an interpretation of  the multiplicity of peaks in $C(T)$: they  may indicate a region of meta-stable states where  the system undergoes  a reversal of large domains. This behaviour has also been found in~\cite{ref5}.

The very existence of these metastable states, which are absent for a low value of $h_0$, supports the interpretation of the transition as first order. This situation is often found in standard Monte Carlo studies of systems undergoing first order transitions, where the dynamics is based on the canonical ensemble and the system risks to be trapped in a local minimum of the free energy. In these cases it is commonly observed that in a finite temperature region, there are jumps in the order parameter curves as a function of the temperature, with hysteresis in field cooling and field heating loops. Noteworthly, a trace of this situation if found here even though the dynamics of the W-L algorithm doesn't suffer from metastabilities.

\begin{figure}[tbp]
\begin{center}
\includegraphics [width=9.0cm]{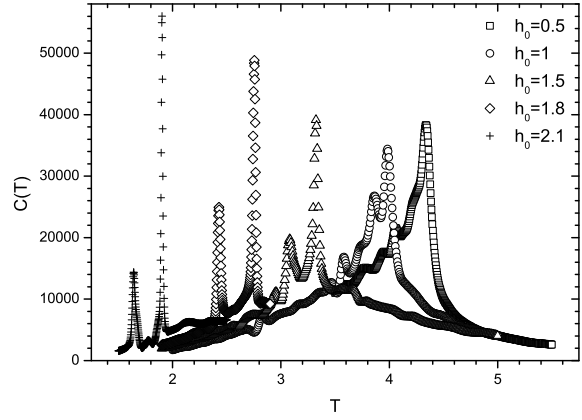}
\caption{$L=24$, $h_0=0.5-2.1$. Specific heat as a function of the
temperature for a single sample and different values of $h_0$. The curve clearly transforms
from a single-peaked $C(T)$ curve characteristic of continuous transitions
to a multi-peak curve where the position and multiplicity of the peaks are
sample dependent}
\label{field_study}
\end{center}
\vspace{1.0cm}
\end{figure}

Figure~\ref{field_study} shows the results of the study of a given sample
for different intensities of the quenched disorder  $h_0$. 
The behaviour clearly changes with the intensity of the random field: while
 $C(T)$ curves  have only one
peak for low field values,  a second one appears,  as the field increases.  
For high field the $C(T)$ curves show clearly more than one peak.

It must be stressed that, for high field values, the double peak in $P_{T^*}(E)$  is always present
for all studied samples of sufficiently large sizes  ( $L \ge 16$) . The position of the peaks may vary as they correspond to different $T^*$  (recall that $T^*$ is  sample dependent).

\section{Discussion}

We have performed a WL study of the 3D-RFIM with a bimodal distribution of
the random fields. Our results clearly show that the critical behavior of
the system depends on the value of $h_0$. For low values of $h_0$ the
transition has all the characteristics of a continuous one: we observed
one peak in $C(T)$ curves at a critical temperature $T_c(L)$. The
corresponding probability distribution of the energy at that temperature is
single-peaked.

For high $h_0$ several indicators of a first order transition are
observed. First $C(T)$ curves become irregular giving rise to
secondary peaks when the field increases. These peaks have already
been observed for the gaussian model and they have been related to
the reversal of large spin domains. This is a phenomenon commonly
encountered in standard Monte Carlo simulations of a magnetic
systems undergoing first order transitions. The system goes through
different meta-stable states when the conditions of the simulation
are modified (typically in field heating and field cooling
simulations). This gives rise to
 hysteresis loops. The passage from one meta-stable
state to the other is done by the reversal of a large domain.

We have also found that the probability energy distribution at the
transition temperature in this high field region is double-peaked, showing
a coexistence of states of the same probability at a given temperature.

This work confirms the results reported in a previous article by one
us~\cite{ref12}. It is interesting to remark that the same result is
now obtained by a completely different calculation method. In fact,
in~\cite{ref12} the bimodal RFIM has been studied using a
combination of standard Monte Carlo (Metropolis) and histogram Monte
Carlo simulations. Both simulations were  performed in the canonical
ensemble. As it is well known, simulations in the canonical ensemble
may be affected by metastabilities near a first order phase
transition as the system risks to be trapped in a local minimum of
the free energy, due to the high energy barriers. In
Ref.\cite{ref12} for a given value of $h_0$, the energy probability
distribution at the transition temperature was found to be double or
single-peaked depending on the particular path the system followed
in the phase space during its evolution.

In the present work, the states of the system are not sampled using the
canonical ensemble. The WL algorithm gives an estimate of the density of
states by performing a random walk on the energy space, so the system should
not be trapped in a metastable state.

We also observe large sample to sample fluctuations in the location and the
height of the specific heat maxima in agreement with Malakis et
al.~\cite{ref14}. At a fixed size, each sample has its own critical temperature
and even the number of secondary peaks of the specific heat depends on the
sample. This implies that the average over disorder of any quantity $A(T)$
in order to locate the critical temperature for a given size, $T_c(L)$ is
not meaningful.  Moreover, by doing so all information concerning an
eventual first order transition would be washed out. This has been the
working method in previous works performing sample averaging, ie:~\cite{ref2} and
it is perhaps one of the reasons for the remaining controversy.

Malakis et al.~\cite{ref14} pointed out the need to take into account the
sample to sample fluctuations when performing averages over the quenched
disorder. Nevertheless, in their work each sample has been studied using the
technique of critical minimum energy subspace described in~\cite{cemes}.
The method assumes a continuous transition. In fact it is based on the hypothesis
of a gaussian-like energy distribution at the pseudo critical temperature
and on the validity of second order finite size scaling relationships. These hypothesis are not valid when $P_T(E)$ is double peaked.

\end{document}